\documentclass[aps,prl,twocolumn,superscriptaddress,preprintnumbers,nobibnotes]{revtex4-1}

\usepackage{graphicx}  
\usepackage{dcolumn}   
\usepackage{bm}        
\usepackage[normalem]{ulem}
\usepackage[usenames,dvipsnames]{xcolor}
\usepackage{amssymb,amsmath,amsfonts}   
\usepackage{siunitx}
\usepackage{hyperref}

\newcommand*{\SII}[2]{\SI[parse-numbers=false]{#1}{#2}}
\DeclareSIUnit\basepair{bp}
\newcommand*{\fig}[1]{Fig.~\ref{#1}}
\newcommand*{\figs}[1]{Figs.~\ref{#1}}
\newcommand*{\eq}[1]{Eq.~\ref{#1}}
\newcommand*{\sctn}[1]{Section~\ref{#1}}

\newcommand*{\kbt}{k_\textmd{B}T}
\newcommand*{\wi}{\textmd{Wi}}
\newcommand*{\shearrate}{\dot{\gamma}}
\newcommand*{\rgo}{R_{g0}}
\newcommand*{\tb}{\tau_{b}}
\newcommand*{\txi}{\tau_{\xi}}

\begin{document}

\preprint{to be published in \textit{Physical Review X}}

\title{Rotation-induced macromolecular spooling of DNA}

\author{Tyler N. Shendruk\footnote{T.N.S., D.S. and D.B. contributed equally to this work. }}
\email[Corresponding author: ]{tshendruk@rockefeller.edu}
\affiliation{The Rockefeller University, 1230 York Avenue, New York, New York, 10065}
\affiliation{The Rudolf Peierls Centre for Theoretical Physics, University of Oxford, 1 Keble Road, Oxford, OX1 3NP, United Kingdom}

\author{David Sean\textsuperscript{*}}
\affiliation{Department of Physics, University of Ottawa, 150 Louis-Pasteur, Ottawa, K1N 6N5, Canada}

\author{Daniel J. Berard\textsuperscript{*}}
\affiliation{Department of Physics, McGill University, Montreal, H3A 2T8, Canada}

\author{Julian Wolf}
\affiliation{Department of Physics, McGill University, Montreal, H3A 2T8, Canada}

\author{Justin Dragoman}
\affiliation{Department of Physics, McGill University, Montreal, H3A 2T8, Canada}

\author{Sophie Battat}
\affiliation{Department of Physics, McGill University, Montreal, H3A 2T8, Canada}

\author{Gary W. Slater}
\affiliation{Department of Physics, University of Ottawa, 150 Louis-Pasteur, Ottawa, K1N 6N5, Canada}

\author{Sabrina R. Leslie}
\email[Corresponding author: ]{sabrina.leslie@mcgill.ca}
\affiliation{Department of Physics, McGill University, Montreal, H3A 2T8, Canada}

\date{\today}

\begin{abstract}
Genetic information is stored in a linear sequence of base-pairs; however, thermal fluctuations and complex DNA conformations such as folds and loops make it challenging to order genomic material for \textit{in vitro} analysis.
In this work, we discover that rotation-induced macromolecular spooling of DNA around a rotating microwire can monotonically order genomic bases, overcoming this challenge.
We use single-molecule fluorescence microscopy to directly visualize long DNA strands deforming and elongating in shear flow near a rotating microwire, in agreement with numerical simulations.
While untethered DNA is observed to elongate substantially, in agreement with our theory and numerical simulations, strong extension of DNA becomes possible by introducing tethering.
For the case of tethered polymers, we show that increasing the rotation rate can deterministically spool a substantial portion of the chain into a fully stretched, single-file conformation. 
When applied to DNA, the fraction of genetic information sequentially ordered on the microwire surface will increase with the contour length, despite the increased entropy. 
This ability to handle long strands of DNA is in contrast to modern DNA sample preparation technologies for sequencing and mapping, which are typically restricted to comparatively short strands resulting in challenges in reconstructing the genome. Thus, in addition to discovering new rotation-induced macromolecular dynamics, this work inspires new approaches to handling genomic-length DNA strands.
\end{abstract}


\maketitle

\section{Introduction}
Although the double-helix structure of DNA stores genetic information linearly as a sequence of bases at the molecular level, entropy randomizes the three-dimensional conformations of DNA polymers in free solution, making accessing genomic information from long polymers exceedingly challenging. 
To manipulate and study DNA, physical approaches have been developed to spatially arrange DNA in a controlled format, including magnetic and optical traps~\cite{bustamante00,bustamante03}, microfluidic approaches~\cite{renner15,jendrejack04,usta06,jo09}, and nanoconfinement approaches~\cite{persson10}, such as Convex Lens-induced Confinement (CLiC)~\cite{Berard2014} or nanopore confinement~\cite{fyta2015}.
However, \textit{in vivo} manipulation of genomic-length biopolymers remains an outstanding challenge that, with existing approaches, becomes more difficult with increasing strand length.

In this work, we untangle and order individual DNA molecules in long robust sections.
Using fluorescence imaging, we demonstrate that DNA strands deform and elongate in shear flow near a rotating microwire when the rotation rate exceeds a critical value.
While single strands of untethered DNA are observed to elongate substantially in agreement with scaling theory and simulations, full DNA extension becomes obtainable by combining tethering with this DNA spooling approach. We find that
tethering one end of the DNA to the rotating microwire enhances extension, and produces novel ``shofar'' conformations.
By slowly increasing the experimentally realizable rotation rate above an additional critical value, our simulations show a substantial portion of the tethered DNA is spooled into a strongly stretched, single-file curvilinear conformation.
Since the conformation consists of a ``shofar''-type tail and a base-ordered stem, we refer to this as a ``French-horn'' conformation.
In this rotation-induced macromolecular spooling, the fraction of the polymer in the fully ordered stem of the French-horn conformation is found to increase with strand length.

\begin{figure}[!tb]
    \centering
    \includegraphics{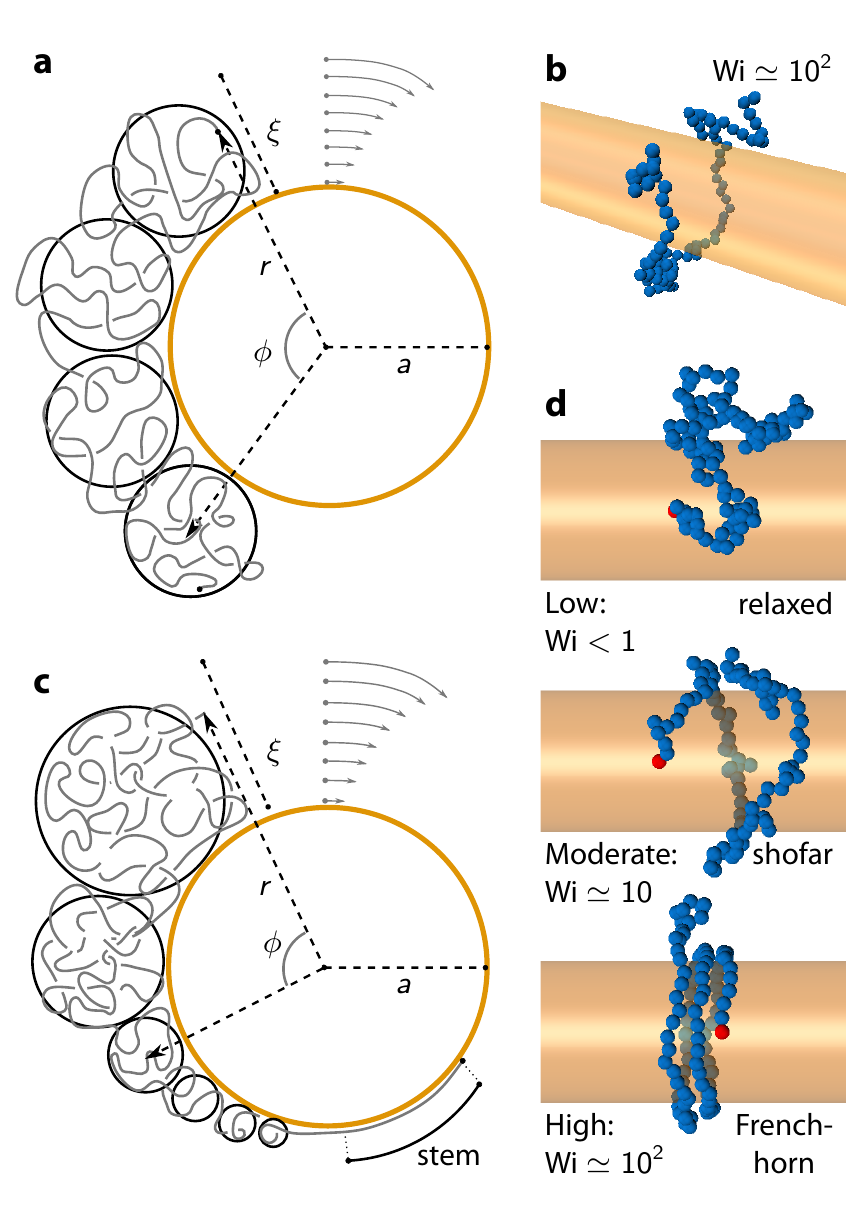}
    \caption{
        \textbf{a.}~%
        The untethered DNA system:
        The rotating cylinder of radius $a$ generates a flow profile $v_\phi\left(r\right)$, which shears the DNA and deforms it. 
        A model for this is a procession of identical blobs of size $\xi$.
        \textbf{b.}~%
        Simulation snapshot portraying a typical steady-state conformation of an untethered DNA chain.
        \textbf{c.}~%
        The tethered system with the French-horn DNA conformation:
        The blob size $\xi$ is a function of angle $\phi$ from the untethered end.
        For sufficiently large Weissenberg numbers $\wi$, a portion of the DNA is fully stretched into a stem section, while the rest is in a horn conformation.
        For smaller $\wi$ there is no stem and the polymer is in the shofar conformation.
        \textbf{d.}~%
        Simulation snapshots portraying typical steady-state tethered conformations.
        Increasing $\wi$ changes the state from relaxed, to shofar, to French-horn conformations.
        Sufficiently long, rapidly rotated chains wrap around the circumference forming a rim of strongly stretched DNA. 
    }
    \label{fig:schematic-theory}
\end{figure}

\subsection{Rotation-induced Spooling}

The physical mechanism for compactly wrapping stretched DNA around a spinning microwire is presented in \fig{fig:schematic-theory}a for untethered DNA.
The rotation drives an azimuthal flow, which shears the deformable DNA and draws it towards the surface of the microwire.
When the DNA is untethered, the biopolymer deforms and elongates.
While the rotation rates required to strongly stretch untethered DNA are very high moderate rotation rates are sufficient to strongly stretch significant portions of the DNA when one end is tethered to the rotating microwire (\figs{fig:schematic-theory}c and~\ref{fig:schematic-theory}d).
The stretched DNA is tautly wrapped and unraveled into a coordinated single-file conformation.
By gradually increasing the rotation rate, the fraction of segments in the strongly stretched stem is increased, while chain-breaking events can be avoided and the excluded volume interactions between the blobs ensure that overlaps are rare.
In this way, substantial portions of single molecules of sufficiently long DNA can be organized into single-file conformations, deterministically ordering the genetic information on a cylindrical surface.

In prior work using planar geometries, DNA and other long macromolecules have been demonstrated to deform when subjected to experimentally achievable flows~\cite{perkins97,saito2011}.
For example, a wall-tethered flexible polymer subject to sufficient shear rates deforms into a string of blobs~\cite{ladoux00,gratton05}.
The physical mechanisms leveraged in this study can be understood by considering the deformation of untethered DNA.
We thus consider untethered strands, after first introducing our methods, before subsequently analyzing tethered chains.

\section{Methods}

When a cylindrical microwire of radius $a$ is rotated with rate $\Omega$, the no-slip boundary condition generates a flow profile $ v_{\phi}\left(r\right) = \Omega a^2/r$ and a rapidly decaying shear rate $\shearrate\left(r\right) = -\Omega\left(a/r\right)^2$ at a distance $r$ from the centre of the cylinder (\fig{fig:schematic-theory}).
Each strand of DNA is composed of $N$ Kuhn segments of length $b \approx \SI{100}{\nano\meter}$ and characteristic relaxation time $\tb = \eta b^3 / \kbt \approx \SII{2.4 \times 10^{-4}}{\second}$ in a solvent with dynamic viscosity $\eta$ and thermal energy $\kbt$.
For a chain in a good solvent with a Flory exponent $\nu \approx 3/5$, unperturbed DNA has a relaxation time $\tau \simeq \tb N^{3\nu}$ and a corresponding undeformed radius of gyration $\rgo \simeq b N^\nu$.

\begin{figure}[!tb]
    \centering
    \includegraphics{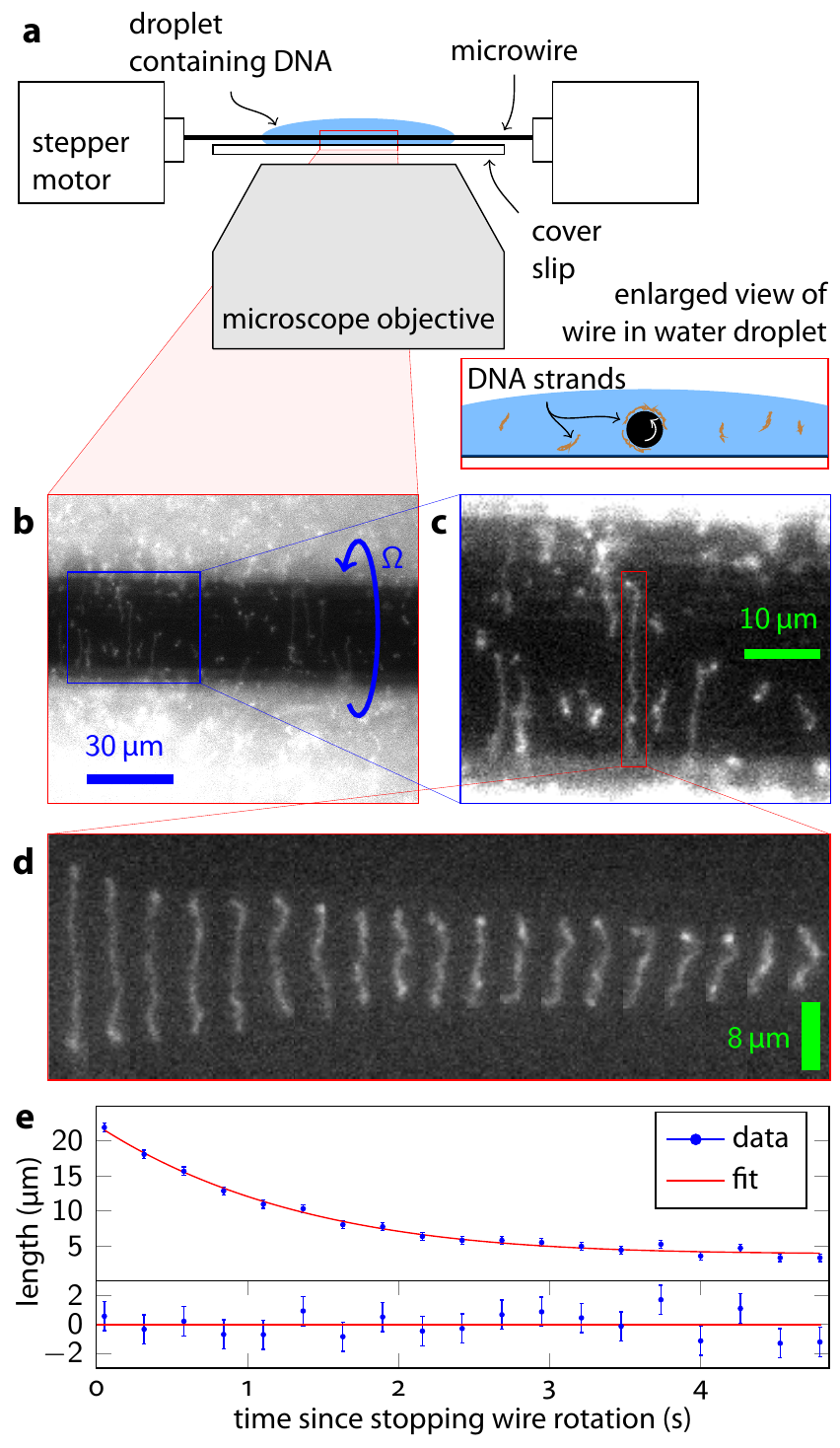}
    \caption{
        The apparatus for untethered strands.
        \textbf{a.}~%
        Schematic representation of the apparatus, described in detail in the text.
        \textbf{b.}~%
        Sample image of a $\SI{50}{\micro\meter}$-diameter tungsten wire (dark horizontal line) rotating in a dilute solution of T4-phage DNA ($\SI{166}{\kilo\basepair}$).
        The contour length of YOYO-1 stained T4-DNA is $\SI[separate-uncertainty = true]{65(2)}{\micro\meter}$.
        Strands of DNA are drawn towards the wire and elongated by the shear flow.
        \textbf{c.}~%
        Image of the microwire $\SI{1}{\second}$ after it has stopped rotating:
        DNA strands have begun to relax to more entropic states.
        \textbf{d.}~%
        Conformational relaxation of a single DNA strand in the period of time immediately after the microwire has stopped rotating (see Movie S1).
        \textbf{e.}~%
        Quantitative measurements of the extension of a single DNA strand during relaxation, along with the fit and associated studentized residuals, used to determine the fully extended length of the polymer ($t=0$).
    }
    \label{fig:sample-data}
\end{figure}

By suspending a microwire directly above a coverslip (\fig{fig:sample-data}a), submerging it in a drop of solution containing DNA, and gradually increasing its rotation rate, we experimentally observe the dynamics of DNA in a rotationally-induced shear flow.
Far from the slowly rotating wire, the shear is small enough that untethered DNA appears relaxed and diffusive.
Nearer to the microwire, the flow dominates and the DNA is advected.
Some polymers, which are initially distributed evenly throughout the solution, are drawn towards the rotating microwire~\cite{Balin2017}.
This radial migration across streamlines towards the microwire is expected to arise from the combined effects of the hydrodynamic interactions with the wire, the nonhomogeneous flow, and the decrease of diffusivity with stretching~\cite{graham11}.
Within the advection regime, many strands are experimentally observed to be elongated by the shear flow (\fig{fig:sample-data}b).
The Weissenberg number $\wi = \shearrate \tau$ expresses the competition between the characteristic shear forces and the polymer's intrinsic relaxation time $\tau$.
The shear deforms the DNA when $\wi \gtrsim 1$ (\fig{fig:sample-data}b and c).
By substituting in the relaxation time and shear rate, we predict that deformation occurs for $r/a \lesssim \left(\Omega \tb \right)^{1/2}N^{3\nu/2}$.
This hints at the benefit of our technique when dealing with long or potentially even genomic-length DNA, since the minimum rotation rate required to achieve deformation is $\Omega_{\wi}^* \tb \simeq N^{-3\nu} $, which drops rapidly with DNA length.

\subsection{Experimental Apparatus}
\label{sec:device}

\begin{figure}[tb]
    \centering
    \includegraphics[width=0.475\textwidth]{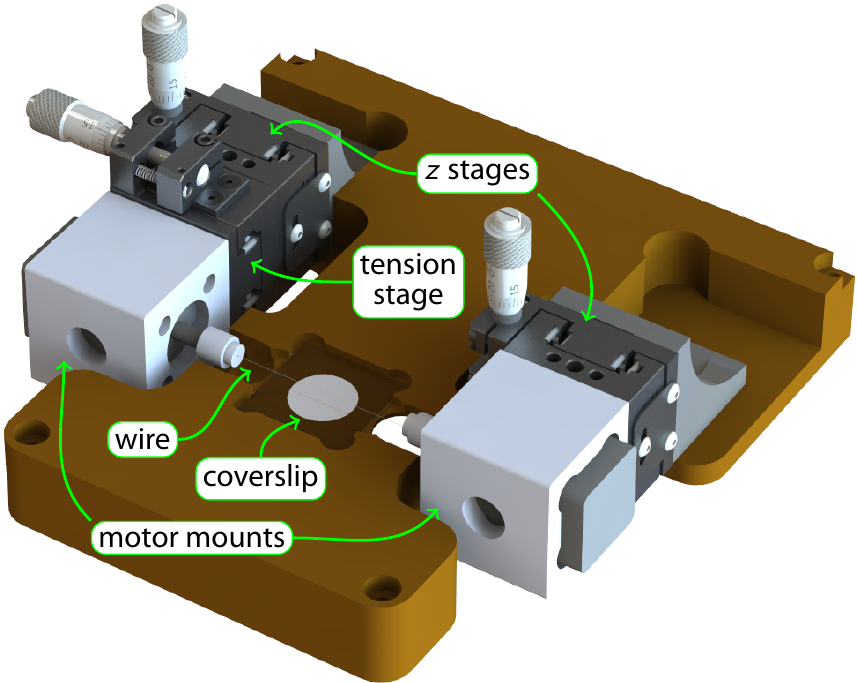}
    \caption{
        The apparatus used to collect data on the rotation-induced macromolecular spooling of DNA.
        The microwire is situated directly above the cover slip using the $z$ stages of the two motor mounts and the tension stage is adjusted so that the microwire is taut.
        Solution containing DNA is dropped onto the cover slip such that a portion of the wire is submerged.
        The motors are driven simultaneously by an Arduino microcontroller.
    }
    \label{fig:apparatus}
\end{figure}

The apparatus used to perform the experiments is shown in \fig{fig:apparatus}.
A $\SI{50}{\micro\meter}$  diameter tungsten wire (Malin co.) is held and spun at both ends by two stepper motors (Pololu Robotics \& Electronics part 1204).
Each end of the wire is held within a short piece of fluorinated ethylene propylene (FEP) microfludic tubing with an inner diameter of $\SI{50}{\micro\meter}$ (IDEX Health \& Science), each of which is mounted to one of the stepper motors using couplers machined from aluminum.
The small inner diameter of the tubing serves to reduce microwire wobble.
The mounting points of the wires do not exactly align, which results in minor wobbling about the rotation axis.

Each stepper motor is mounted on a $z$-axis-stage (Thorlabs part MS1S) which allows its height to be varied.
This allows the vertical position of the wire above the sample coverslip, as well as the angle of the wire to the horizontal, to be set precisely.
One of the stepper motors is also mounted on a horizontal stage which allows the distance between the motors to be varied, so that a controlled tension can be applied to the microwire.
The motors and stages are mounted to a base machined from aluminum.
The rotation of the motors is controlled by an Arduino microcontroller board.
The current maximum rotation rate is not limited by the motors but rather by vibrations and wobbling that hinder accurate imaging of the DNA. 

The apparatus mounts on an inverted fluorescence microscope (Nikon Ti-E), and experiments are performed with a $60 \times$ NA $1.0$ water immersion objective (Nikon CFI Plan Apo VC 60XWI) and an Andor iXon3 EMCCD camera at an EM gain of $270$.
We estimate the depth-of-field of our optical system to be about $\SI{1.2}{\micro\meter}$. This means that, due to the curvature of the microwire, the entire length of an extended DNA molecule cannot be in perfect focus at once. The longest segment of DNA which can lie within the focal plane at once is about $\SI{16}{\micro\meter}$ for a $\SI{50}{\micro\meter}$-diameter wire, so some sections of the measured T4-DNA molecules are inevitably slightly out of focus. 
In order to observe the DNA, the YOYO-1 dye (\sctn{sec:DNA-prep}) is excited using $\SI{1.5}{\milli\watt}$ from a $\SI{488}{\nano\meter}$ Coherent Sapphire laser.

\subsection{DNA Preparation}
\label{sec:DNA-prep}

For untethered experiments, T4-phage DNA ($\SI{166}{\kilo\basepair}$, New England Biosciences) is used at a concentration of $\SI{0.1}{\micro\gram/\milli\liter}$.
The DNA are stained with YOYO-1 fluorescent dye (Life Technologies) at a ratio of $1$ dye molecule per $10$ base-pairs.
YOYO-1 increases the DNA contour length at this staining ratio from $\SI{56.4}{\micro\meter}$ to $\SI{65 \pm 2}{\micro\meter}$ for T4-DNA~\cite{Gunther10}.
Experiments are carried out in $0.5 \times$~tris-borate-EDTA (TBE) buffer, with $\SI{3}{\%}$ $\beta$-mercaptoethanol (BME) added as an anti-photobleaching agent.
Before beginning untethered experiments, the microwire is first passivated using a solution of $\SI{10}{\%}$ $\SI{55}{\kilo\dalton}$ polyvinylpyrrolidone (PVP) in 0.5X TBE to reduce sticking of DNA to the glass coverslip and tungsten microwire.

For a solvent viscosity similar to that of water, the bulk relaxation time and radius of gyration of T4-DNA are $\tau = \SI{0.74}{\second}$ and $R_{g0} = \SI{1.46}{\micro\meter}$~\cite{Balducci06} with approximately $N\approx500$ Kuhn segments. 
The Flory exponent for T4-DNA in $0.5 \times$~TBE (a good solvent) is $\nu = 0.558$, which is intermediate between a Gaussian chain and a swollen excluded volume chain to account for the finite length and width of the double-stranded DNA~\cite{Tree13}.

\subsection{Measuring Extension}
\label{sec:DNA-extension}

To accurately measure the end-to-end lengths $L$ of the untethered DNA in the experimental apparatus, the relaxation from the extended state is fit.
The microwire rotates, extending the near-wire polymers, and then the rotation is suddenly halted.
The DNA strands relax to a more highly entropic state with decreased extension and the initial length $L(t = 0)$ is extracted is obtained from~\cite{klepinger2015}
\begin{equation} \label{eq:strand-relaxation}
    L^2\left(t\right) - 4 \rgo^2 = A \,N b \,e^{-t / \tau},
\end{equation}
where $t$ is the time since the rotation of the microwire has stopped and $A$ is a proportionality constant that is allowed to vary when fitting.
To estimate the extension when the wire is rotating, this function is fit to the relaxation of each polymer, and the initialextended length when the wire is still rotating $L(t = 0) = \sqrt{4 \rgo^2 + A \,N b}$ is extracted from the fit (\fig{fig:sample-data}e).
The fit to \eq{eq:strand-relaxation} is found to be accurate for the range of rotation rates utilized in this study. 

Each individual measurement of the elongation length of the DNA strand at time $t$ requires locating the positions of its two ends.
The uncertainty in the experimental measurement of each of these positions is dominated by the pixel size of the camera, corresponding to $\SI{267}{\nano\meter}$.
The uncertainty in each measurement is then $\delta L = \SI{533}{\nano\meter}$, while the uncertainty in the fully extended length of the polymer is extracted from the fit to its relaxation.
There is large variance in the extent of elongation as expected from results on tumbling DNA in simple shears~\cite{Teixeira05,Schroeder05a}.

\subsection{DPD Algorithm}
\label{sec:dpd}
The rotation-induced macromolecular spooling does not rely on chemical details of DNA---it is a non-specific physical phenomenon.
For this reason, a coarse-grain Dissipative Particle Dynamics (DPD) numerical model is used to simulate generic DNA molecules to extend our understanding of DNA dynamics in three dimensional azimuthal flows at various rotation rates with hydrodynamic interactions.

Both DNA monomers and fluid particles are represented by soft particles of mass $m$ and size $\sigma$.
The $i^\textmd{th}$ particle is described by its coordinates in continuous-space, which evolve in response to inter-particle forces with other particles (labeled $j$).
In DPD, these are commonly pair-wise conservative forces $\vec{F}_{ij}^C$ with a soft repulsion with a cut off $r_c$ and strength $a_{ij}$, dissipative forces $\vec{F}_{ij}^D$ with inter-particle frictional drag $\gamma_\textmd{DPD}$, and random forces $\vec{F}_{ij}^R$ due to the thermal energy $\kbt$.
An extended discussion of these forces can be found in Ref.~\cite{slater2009modeling}.

The number density of DPD beads in the simulations is $\rho_f$ and through the fluctuation-dissipation theorem the drag is $\gamma_\textmd{DPD}=\sigma^2/\left(2\kbt\right)$.
The simulation units are $[m=1; r_c=1; \kbt=1]$ and we choose $\sigma=3$, $\gamma_\textmd{DPD}=4.5$, $\Delta t=0.01$, $\rho_f=3$, and $a_{ij}=25$.

The DPD beads that form the DNA strand have an additional conservative bonding force to sequentially polymerize $N=100$ beads.
For this, we opt for the widely used FENE spring potential  $U_{\textrm{FENE}}$, in conjuction with the purely repulsive truncated Lennard-Jones potential $U_\mathrm{WCA}$~\cite{slater2009modeling}.
We note that the both $U_{\textrm{FENE}}$ and $U_\mathrm{WCA}$ are only applied between consecutive coarse-grained beads.
The effective coarse-grained monomer size is determined from the $U_\mathrm{WCA}$ and chosen to be the same as the DPD beads $r_c=1$.
The DPD repulsion suffices to model long-ranged excluded-volume interactions, which yield the expected size exponent for a real polymer in a  good solvent~\cite{Jiang07}.
This allows us to obtain a monomer relaxation time  $1<\tb<0.1$~\cite{Jiang07}, which we approximate as $\tb\approx0.3$.
The presence of the DPD fluid ensures that hydrodynamic interactions between monomers are reproduced, leading to Zimm dynamics.
The bulk relaxation time of DPD chains kept in the vicinity of the surface of the microwire by tethering in quiescent solvent is measured to be $\tau=920$, corresponding to $\tau = \SI{0.74}{\second}$ when modelling T4-DNA.
Likewise, each DPD bead represents roughly ten Kuhn segments of T4-DNA, since the DPD chain consists of $N=100$ beads, while T4-DNA is composed of $N\sim10^3$ Kuhn segments. 
Such long DNA can be rescaled and coarse-grained as a freely-jointed chain on scales above the DPD bead size, while dynamics on smaller length scales are not modelled. 

We numerically model the microwire as a semi-infinite cylindrical surface of radius $a=5$ composed of three layers of DPD beads, which rotate with a fixed angular velocity.
The rotation of these wall beads produces the shear rate $\shearrate$ that deforms the polymer, which hydrodynamically interacts with the cylinder surface through the DPD solvent. 
Relative to $\rgo$, the simulated cylinder is smaller than the experimentally employed tungsten wire, which may impact cross-streamline migration. 
A concentric cylinder of radius $a_\textmd{out}=35$ provides a finite size to the system and is constituted by a single layer of DPD beads that rotates with the theoretically expected speed $v_\phi\left(r=a_\textmd{out}\right)=\Omega a^2/a_\textmd{out}$.
The resulting shears match the theoretical expectations except very near the microwire surface, where minor packing and slip-effects cause the flow profile to be over predicted.
We apply periodic boundary conditions along the microwire's axial direction with a system thickness of $15$.
We observe that the soft nature of the DPD beads can lead to slight polymer penetration into the inner rod when the rotation rate is sufficiently high.
To reduce this effect, we add a smooth mathematical cylindrical surface that is transparent to the solvent beads (hence does not explicitly affect the flow) but unto which we apply $U_\mathrm{WCA}$ for every polymer bead.
With the surface characterized by a radius $a-1/2$, the polymer feels a steric repulsion and does not penetrate the surface of the nominal rod.

The DNA model is initialized tightly wrapped around the cylinder and left to equilibrate.
By noting the monomer position in a cylindrical coordinate system $\vec{r}_i(r_i,\phi_i,z_i)$, it is possible to find the instantaneous span $L_s \equiv \langle r \rangle \Delta \phi=\langle r \rangle \left( \phi_\mathrm{max}-\phi_\mathrm{min} \right)$.
A tightly wrapped polymer is thus initialized at a high value of $L_s$, which exponentially decreases to a steady state value (characterized by $\tau_\mathrm{exp}$).
An exponential fit to \eq{eq:strand-relaxation} is conducted to determine the steady-state curvilinear extension $L$.
As in experiments, only DNA strands in the elongated phase of the tumbling cycle~\cite{Schroeder05a} are analyzed.

Although our simulations demonstrate that the monomers are driven towards the cylinder surface as expected, when the rotation rate is low, untethered polymers can sometimes escape into the bulk. 
This may be due to the fact that the simulated cylinder is effectively smaller than than in experiments, which may affect radial migration. 
Although re-capture events are observed, escaped polymers are not analyzed.
As in simple shear, tumbling dynamics~\cite{Schroeder05b} and U-turn conformations~\cite{Harasim13} are observed in rotation-induced macromolecular spooling.
This sets a lower bound to the rotation rate since the model DNA commonly escapes from the surface of the rotating wire at slow rotation rates.

\section{Untethered DNA}

\subsection{Weakly Extended Untethered DNA}
\label{sec:untethered-weak}

The curvilinear extension $L$ of elongated polymers are experimentally measured at a number of rotation rates (\fig{fig:untethered-results}).
Below $\Omega \approx \SI{7}{rpm}$, some strands are drawn towards the rotating microwire; however, their elongation is weak.
In this weak elongation regime ($\wi \lesssim 1$), the DNA behaves as a thermal spring subjected to hydrodynamic stretching forces.
The incident shear rate around a rotating cylinder is expected to be $\shearrate = \Omega a^2/\left(a+\xi\right)^2 \simeq \Omega$, which enacts a characteristic dimensionless drag $F b/\kbt \simeq \Omega\tb N^{2\nu}$ on the weakly deformed DNA. 
The drag force on T4-DNA at the microwire surface is roughly $F \sim \eta\rgo \times\left(\shearrate\rgo\right) \sim \SI{E-2}{\pico\newton}$. 
Since the dimensionless thermal spring force scales as $\simeq L/\left(Nb\right)$, we predict that in the weak elongation regime
\begin{align}
 L/ \left(Nb\right) \simeq \Omega\tb N^{2\nu},
 \label{eq:weakE22}
\end{align}
which can also be written $L/\rgo \simeq \wi N^{1-2\nu}$ and is independent of wire diameter.

The DNA extension is thus predicted to increase rapidly with rotation rate as $L\sim\wi$ in this regime, which is consistent with the low rotation rate data in \fig{fig:untethered-results}.
However, a crossover from an initially rapid rise for $\wi\lesssim1$ to a much slower rise when $\wi>1$ is expected; the experimental data shows a marked crossover from the weak-extension regime to a much lower scaling at moderate rotation rates.

\subsection{Moderately Extended Untethered DNA}
\label{sec:untethered-moderate}

When the rotation rate is moderate ($\wi\gtrsim1$), the DNA deforms and aligns along the azimuthal streamlines at the surface of the rotating cylinder (\fig{fig:sample-data}c-d). 
For a limited regime, this moderate elongation can be roughly regarded as a string of blobs, in which the statistical breadth of the deformed chain (transverse to the flow direction) is described by the characteristic size $\xi=\kbt/F$ (\fig{fig:schematic-theory}b). 
This moderate elongation regime is bordered by the weakly extended regime discussed in \sctn{sec:untethered-weak} for forces that do not deform the conformation ($\kbt/\rgo \sim \SI{3E-3}{\pico\newton}$) and by the upper limit of strong deformation, in which the blob framework breaks down as $\xi$ approaches the Kuhn length ($\kbt/b \sim \SI{6E-2}{\pico\newton}$). 
While the upper limit is a mechanical property of double stranded DNA, the lower limit decreases as $\rgo \sim N^\nu$ increases. 
In this limited moderate extension regime, genomic-length DNA of many Kuhn segments can be considered a procession of blobs, in which tension is equally distributed throughout the chain and each blob has a theoretical relaxation time $\txi \simeq \tb \left(\xi/b\right)^{3}$. 
Since shear and relaxation balance when the Weissenberg number for each is unity, the scaling prediction for the untethered characteristic size scale is 
\begin{align}
 \xi/b &\simeq \left(\Omega\tb\right)^{-1/3} \qquad \textmd{if } \xi \lesssim a.
 \label{eq:untetheredBlob}
\end{align}
This result can be written as $\xi/\rgo \simeq \wi^{-1/3}$.

\begin{figure}[tb]
    \centering
    \includegraphics{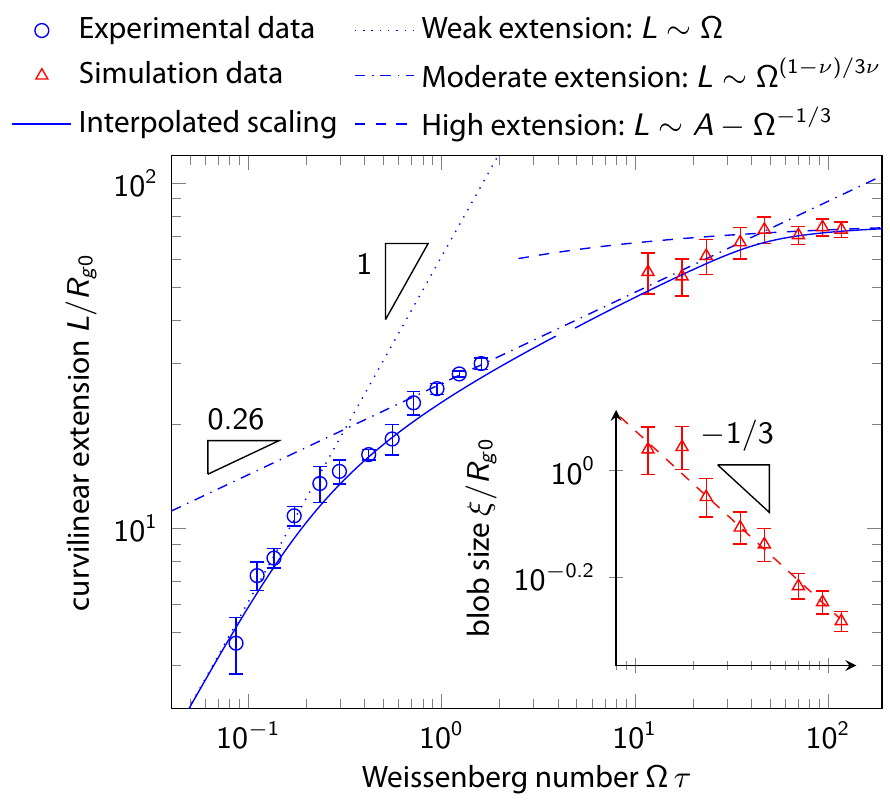}
    \caption{
        Curvilinear end-to-end extension of untethered T4-DNA ($\SI{166}{\kilo\basepair}$) and DPD polymers for varying rotation rates.
        The model-dependent shift between the high Weissenberg-number simulations and experiments is removed by setting the $\wi = 10$ value to the fitted strong extension value.
        Dotted lines show the predicted scaling for weak extensions at low $\wi$ (\eq{eq:weakE22}; $L\sim\wi$). 
        Moderate extensions at intermediate $\wi$ are shown as dash-dot lintes (\eq{eq:untetheredE2E}; $L\sim\wi^{\left(1-\nu\right)/3\nu}\sim\wi^{0.26}$), while dashed lines show strong extensions at the highest $\wi$ (\eq{eq:untetheredStrong}). 
        \textbf{Inset.}
        Blob size $\xi$ of the DPD polymer.
        Dashed line shows the scaling predicted by \eq{eq:untetheredBlob}.
    }
    \label{fig:untethered-results}
\end{figure}

The conformational breadth of the DNA transverse to the flow direction is experimentally inaccessible due to the diffraction limited resolution. 
Therefore, we consult numerical experiments to directly investigate the conformational breadth. 
DPD simulations explicitly estimate size as the average distance of the monomers from the surface of the wire $\xi=\langle r \rangle-a$ (\fig{fig:untethered-results} inset).
For sufficient rotation rates, $\xi$ scales with rotation rate in the manner predicted by \eq{eq:untetheredBlob}.
At low rotation rates, the DNA often diffuses into bulk solution from the wrapped state, placing a lower bound on the rotation rates that can be numerically investigated (see \sctn{sec:dpd}).

The curvilinear end-to-end distance of the highly deformed DNA on the wire surface is predicted to be
\begin{align}
 L/ \left(Nb\right) &\simeq \left( \Omega\tb \right)^{\left(1-\nu\right)/\left(3\nu\right)}
 \label{eq:untetheredE2E}
\end{align}
by considering the number of Kuhn segments in a single blob and the total number of blobs.
This result is equivalent to $L/\rgo \simeq \wi^{\left(1-\nu\right)/\left(3\nu\right)}$.
The extension increases linearly with contour length $Nb$ such that that longer chains stretch more easily than shorter chains in this moderate extension regime, but the extension is only weakly dependent on rotation rate in this moderately extended regime.
Inserting $\nu=0.558$ for T4-DNA predicts a weak dependence on rotation rate ($L/\rgo \sim \wi^{0.26}$) for this regime, which stands in contrast to the initial linear dependence in \eq{eq:weakE22} for weakly stretched chains. 
Though the effective Flory exponent may vary slightly for blobs consisting of fewer Kuhn lengths, since double stranded DNA is neither sufficiently thin nor stiff to be a Gaussian chain ($\nu=1/2$) the double stranded minimum value ($\nu=0.535$~\cite{Tree13}) is only 4\% less than the full chain value for T4-DNA ($\nu=0.558$~\cite{Tree13}). 

At the highest experimentally achievable rotation rates in \fig{fig:untethered-results}, the extension of the untethered DNA scales in accordance with \eq{eq:untetheredE2E}.
Due to limitations on the current apparatus, the microwire cannot be steadily driven above $\Omega \approx \SI{130}{rpm}$ without inducing significant wire wobble.
DPD simulations further support this scaling at moderate rotation rates. 
The coarse-grained simulations replicate the experimental scaling in the moderate extension regime described by \eq{eq:untetheredE2E}. 

Although the moderate extension regime is evident in both the experimental and numerical untethered results (\fig{fig:untethered-results}), relatively large experimental rotation rates are required to deform the T4-DNA substantially more. 
The DPD data extend to higher $\wi$ and exhibit evidence of a strong stretching limit by deviating from \eq{eq:untetheredE2E} (\fig{fig:untethered-results}). 
In this highly extended regime, the dimensionless restoring force goes as $Fb/\kbt \sim \left[1-L/(Nb)\right]^{-2}$, while the nearness of the wire surface suggests a drag force $\sim \eta \left(Nb\right) \times \left(\shearrate \delta r\right)$ with radial excursions of $\delta r\approx b\left[1-L/(Nb)\right]$. 
Combined, these suggest 
\begin{align}
 1-L/(Nb) &\simeq \left( N\Omega\tb \right)^{-1/3}, 
 \label{eq:untetheredStrong}
\end{align}
which is consistent with the high $\wi$ DPD data. 
In this strong force limit, the coarse-grained DPD simulations are not expected to quantitatively reproduce the semiflexible nature of double stranded DNA; however, we expect the qualitative conclusion that the curvilinear extension slows at high Weissenberg number to hold for double stranded DNA. 
To experimentally obtain such strongly stretched chains would require $\xi\rightarrow b$ and, therefore, $\Omega_\textmd{full}^* \simeq \tb^{-1} \approx \SI{4000}{\per\second}$, which would be experimentally challenging. 
However, by anchoring one end of the DNA to the rotating microwire, a substantial portion of sufficiently long chains can be strongly stretched. 

\section{Tethered DNA}

Untethered strands of DNA readily deform in the vicinity of the rotating microwire, but reaching the fully stretched limit is infeasible and fluctuations due to tumbling dynamics are significant.
However, by anchoring one end of the DNA to the rotating microwire, a substantial portion of sufficiently long chains can be strongly stretched.

For the tethered experiments, $\lambda$-phage DNA ($\SI{48.5}{\kilo\basepair}$, New England Biosciences) is used, also at a concentration of $\SI{0.1}{\micro\gram/\milli\liter}$ and stained with YOYO-1 fluorescent dye as in the untethered case.
Staining increases the $\lambda$-DNA contour length from $\SI{16.5}{\micro\meter}$ to $\SI{19.0 \pm 0.7}{\micro\meter}$~\cite{Gunther10}.
Custom Cy5-labeled oligos complementary to the 12-bp single-stranded overhangs at the ends of $\lambda$-phage DNA are added (IDT).
The hydrophobic character of the Cy5 end-labels results in a significant increase in surface-ahesion.
The glass coverslip is passivated with PVP as in the untethered case, but the microwire is not.
Tethered DNA molecules, for which one end has attached to the wire, are used for quantitative analysis of the tethered case.
Since the hydrophobic Cy5-labels are positioned at the DNA ends, immobilization occurs at the ends.
In DPD simulations, the polymer is tethered at one end via a FENE potential. 

Similar to the classical differences between a chain dragged by one end through a quiescent fluid and a wall-tethered chain subjected to shear~\cite{saito2011,brochardwyart93}, the tethered system setup differs from a chain attached to a filament~\cite{laleman16} in that it is the shearing flow due to rotation that leads to deformation.
At low rotation rates ($\Omega<\Omega_{\wi}^*$), the tethered DNA is expected to remain relaxed and simply rotate with the anchoring point.
Simulation snapshots at low rotation rates (\fig{fig:schematic-theory}d; $\wi\simeq10^{-0.3}$) show that the tethered DNA molecules follow the rotation in a quasi-static manner.
They remain in a thermalized state throughout the complete simulation runs because the shear is too low to stretch the DNA into an out-of-equilibrium conformation.
The critical rotation rate for deformation is predicted to scale as $\Omega_{\wi}^*\tb \simeq N^{-3\nu} $, which is the same as in the untethered state and is confirmed by DPD simulations of tethered DNA (\fig{fig:tethered-results}a).

\subsection{Shofar Conformation}
\label{sec:shofar}

\begin{figure}
    \centering
    \includegraphics[width=0.42\textwidth]{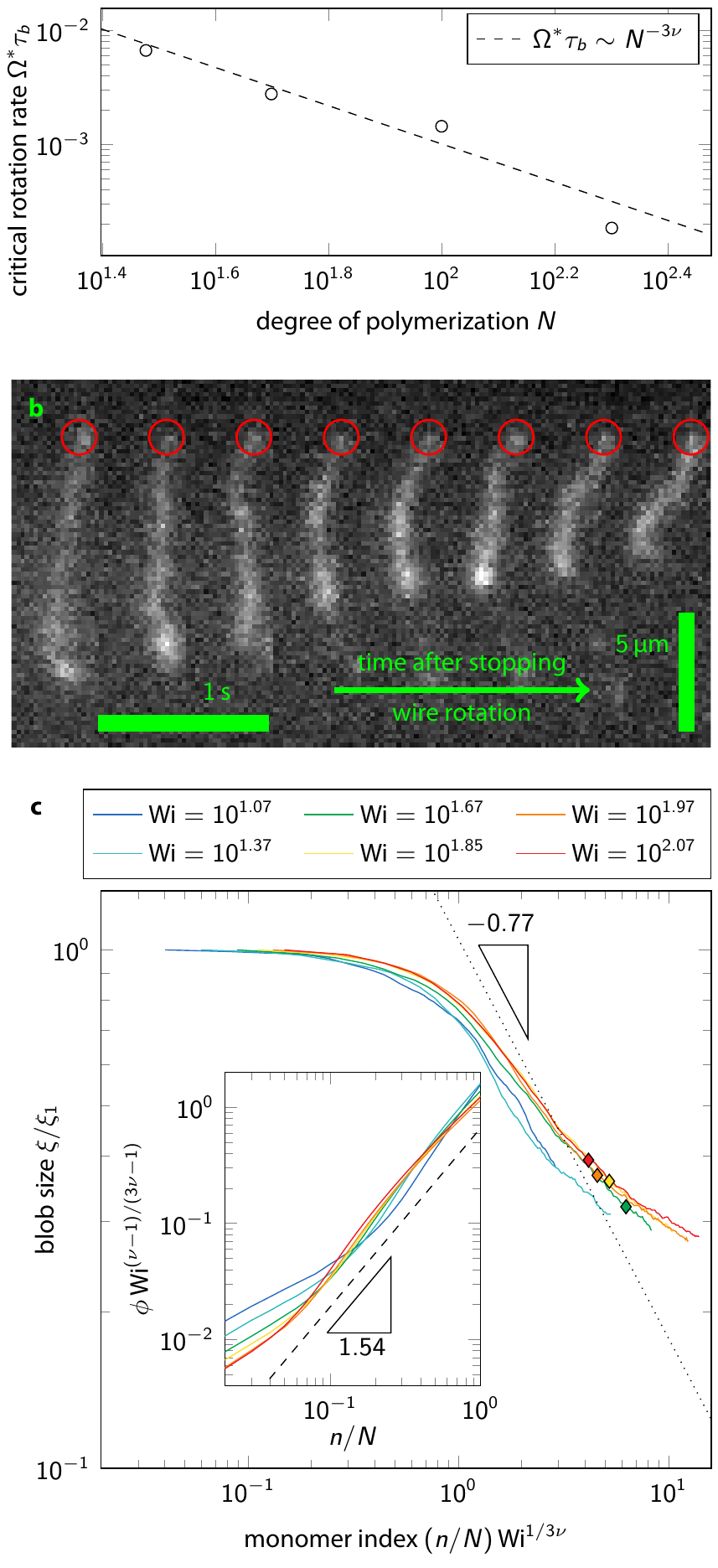}
    \caption{
        Spooling tethered DNA strands.
        \textbf{a.}~%
        Simulations of the critical rotation rate for which deformation of tethered DNA occurs. The solid line shows the predicted scaling of $\Omega_{\wi}^* \sim N^{-3\nu} $. 
        \textbf{b.}~%
        Conformation of a single $\lambda$-phage DNA strand in the period immediately after the microwire has stopped rotating at $\Omega = \SI{130}{rpm}$, showing the relaxation of the polymer (see Supplemental Video 2).
        The end of the polymer that appears towards the top of the image is tethered to the wire.
        The conformational size is observed to increase steadily away from the tethering point, as predicted for a tethered polymer in the shofar configuration. 
        \textbf{c.}~%
        French-horn conformations seen via DPD blob size $\xi$ as a function of segment index from the free-end.
        Various rotation rates are collapsed according to \eq{eq:tetheredIndex} with blob size normalized by the free-end blob size $\xi_1$.
        Dotted line shows scaling prediction $\xi \sim n^{-\nu/\left(3\nu-1\right)} \sim n^{-0.77}$.
        Diamonds ($\diamond$) denote the predicted transition from shofar to stem at $n^* \simeq \left(\tb\Omega\right)^{-1}$.
        \textbf{Inset.}~%
        Wrapping angle from the free end, collapsed via \eq{eq:tetheredAngle}.
        Dashed line shows the scaling $\phi \sim n^{2\nu/\left(3\nu-1\right)} \sim n^{1.54}$.
    }
    \label{fig:tethered-results}
\end{figure}

For greater rotation rates, the tethered polymer deforms into a conformational profile that decreases in size as a function of the angle $\phi$ from the free end (\fig{fig:schematic-theory}c).
At a  rotation rate $\Omega \approx \SI{130}{rpm}$ a configuration of varying transverse conformational breadth is readily achievable and an example of the growth of the conformational size away from the tethering point is qualitatively clear in \fig{fig:tethered-results}b.
If the planar case may be described as a ``trumpet'' conformation~\cite{buguin96} then our experiments produce a cylindrical ``shofar'' conformation. 

DPD snapshots also demonstrate this shofar conformation (\fig{fig:schematic-theory}d), showing that a moderate rotation rate is sufficient to deform a long polymer and wrap it around the microwire. 
There remain visible thermal fluctuations along the backbone, and the conformational size can be seen to increase with the distance to the tethering point, as in the experiments shown in \fig{fig:tethered-results}b. 
The simulations can quantitatively measure the statistical shape of the shofar conformation $\xi\left(n\right)$ as a function of segment index from the free-end (\fig{fig:tethered-results}c). 
A free-end regime is clearly identifiable at small $n$ in numerical measurements of the transverse conformational size $\xi$, while at moderate $n$ the size decreases rapidly until the tethering point is approached. 
Although the range is limited by the contour length of the simulated chain, the mid-region conformation of the simulated chain appears to exhibit a rapid decaying scaling law. 

In order to understand the observed shofar conformation, we extend the scaling theory for highly extended untethered DNA (\eq{eq:untetheredBlob}) to a tethered chain, in which the tension in each Kuhn segment varies with the distance from the free end. 
Temporarily, consider a series of well-defined blobs for a sufficiently long genomic-length polymer: The $i^\textmd{th}$ blob is described by the radial distance $r_i=a+\xi_i$ and arc length $s_i=r_i\phi$ (\fig{fig:schematic-theory}c). 
The drag force on each is $f_i \simeq \eta v_\phi\left(r_i\right) \xi_i$ but the tension on each segment is due to the total drag of all the preceding blobs. 
However, the tension does not vary stepwise through a procession of well-defined blobs but rather varies continuously from segment to segment as $\sum_j f_j = \sum_j \eta v_\phi \xi_j \rightarrow \int \eta v_\phi \xi \left(ds/\xi\right)$, which is still expected to be locally balanced by the thermal force $\kbt/\xi$. 
Therefore, for sufficiently long polymers, the scaling theory predicts that the transverse conformational profile as a function of angle $\phi$ from the free end is
\begin{align}
 \xi\left(\phi\right) / b &\simeq \left( \frac{a}{b} \Omega\tb \phi \right)^{-1/2}.
 \label{eq:tetheredBlob}
\end{align}
This scaling result reproduces the classic planar-shear result when the shear rate $\shearrate$ replaces $\Omega$ and the linear distance replaces the arc length $s = a\phi$~\cite{brochardwyart93,slater04}.

Assuming the tension varies slowly along the chain, we can write the rate of change of the local extension $s$ with segment index $n$ counted from the free end to be $ds/dn \approx b\left( b/\xi \right)^{\left(1-\nu\right)/\nu}$~\cite{brochardwyart93}, or $ds/dn \approx \left(a+\xi\right)\left( d\phi /dn\right)$ by inverting \eq{eq:tetheredBlob}.
Assuming that $a \gg b$ and considering $n$ sufficiently far from the free end, we equate the two forms and integrate to find
\begin{align}
 \phi\left( n \right) \sim \left(b/a\right) \left( \Omega\tb \right)^{\left(1-\nu\right)/\left(3\nu-1\right)} n^{2\nu/\left(3\nu-1\right)}.
 \label{eq:tetheredAngle}
\end{align}
Within the shofar conformation and sufficiently far from the free end, simulations show that the angle from the free end grows with segment index as described by \eq{eq:tetheredAngle} (\fig{fig:tethered-results}c inset; dashed line).
As expected, the predicted scaling does not hold near the free end.

Further details about the conformation can be obtained from the blob size as a function of segment index from the free end.
Substituted into \eq{eq:tetheredBlob}, the angle $\xi\left(n\right)$ produces
\begin{align}
 \xi\left(n\right) / b &\simeq \left( \Omega\tb n \right)^{-\nu/\left(3\nu-1\right)},
 \label{eq:tetheredIndex}
\end{align}
which can be written $\xi\left(n\right) / \rgo \simeq \left( \wi \; n/N \right)^{-\nu/\left(3\nu-1\right)}$. 
As in the planar case, the blob framework (\eq{eq:tetheredIndex}) predicts that the conformational profile diverges as $n \rightarrow 0$.
Therefore, it is truncated at the first blob $\phi_1 \approx \xi_1/a$.
This predicts $\xi_1/b \simeq \left(\Omega\tb\right)^{-1/3} $, which is the same as the untethered conformational size (\eq{eq:untetheredBlob}).
The conformation crosses to the shofar conformation that was observed in the inset and is seen to be well described by \eq{eq:tetheredIndex} (\fig{fig:tethered-results}c; dotted line) for segments in the mid-region of the simulated chain.
Within the blob framework, the conformation at the tethering point is predicted to be $\xi\left(N\right)/b \simeq \left( \Omega\tb N \right)^{-\nu/\left(3\nu-1\right)}$.
However, as the tethered end is approached, the conformational breadth as a function of index begins to saturate to the segment size for many of rotation rates in the simulations (\fig{fig:tethered-results}c).
These curves represent a new conformation.

\subsection{French-horn Conformation}
\label{sec:french-horn}

If the rotation rate is increased to $\Omega_\textmd{horn}^* \tb \simeq N^{-1}$ then the conformational profile at the tethering point is predicted to approach the Kuhn size $b$ and the blob-scaling regime ends.
For higher rotation rates ($\Omega>\Omega_\textmd{horn}^*$), the segments near the tethering point are strongly stretched and the chain adopts a ``French-horn'' conformation with a strongly stretched ``stem'' near the tethering point followed by a shofar-shaped horn conformation as the free end is approached (\fig{fig:schematic-theory}c).
The onset of this French-horn conformation occurs at substantially slower rotation rates than strongly stretched conformations of unthethered polymers.
The segments within the stem and indeed those for which the characteristic blob size is not much larger than the double stranded DNA Kuhn length ($\xi\left(n\right) \not\gg b$) may not be in the scaling regime.
However, sufficiently far from from the stem the conformation of the ``horn'' portion can still be described by blob theory, and the segment index where the transition from shofar to stem occurs is $n^* \simeq \Omega_\text{stem}^* / \Omega \simeq \left(\tb\Omega\right)^{-1}$, and so the fraction of segments in the strongly stretched stem is
\begin{align}
 \lambda &\simeq \frac{N-n^*}{N} = 1 - \left( N\Omega\tb \right)^{-1}.
\end{align}
Since the tension in the chain arises from the cumulated drag of the shofar conformation, obtaining $\lambda\rightarrow1$ requires unphysical rotation rates ($\Omega\rightarrow\Omega_\text{stem}=\tb^{-1}$). Likewise, even strongly stretching 90\% of the chain demands $N\Omega\tb=10$; however, strongly stretching half of the DNA requires a rotation rate that is decreased by nearly an order of magnitude to only $N\Omega\tb=2$. 
Due to limitations on the current apparatus ($\approx \SI{130}{rpm}$ maximum experimental rotation rate), the stem of the French-horn conformation has not yet been experimentally observed. 
Though the relaxation time of a single Kuhn segment ($\tb\sim \SII{10^{-4}}{\second}$) is quite short, $N$ can be extremely large for DNA, such that experimentally obtainable rotation rates are predicted to strongly stretch significant portions of genomic-length biopolymers.

A numerical rotation rate of $\Omega=0.1$ suffices to wrap the majority of the monomers in a single-file manner in simulations (\fig{fig:schematic-theory}d; $\wi \simeq 10^{2}$).
Here, the stem-state dominates with a small proportion of monomers forming a more relaxed conformation as the free end is approached.
Sufficiently long chains encircle the microwire with a strongly stretched and ordered rim of genomic information. 
In simulations, the stem portion does not reach a constant cutoff due to the softness of the DPD beads and slight bond stretching, which results in a reduced but non-zero decrease in effective blob size (\fig{fig:tethered-results}c). 
As in the untethered case, in the strongly stretched stem the DPD simulations provide a qualitative picture but not quantitative predictions for semiflexible double stranded DNA. 
However, the predicted crossover from the shofar to the stem of the French-horn conformation at $n^*$ is seen to approximate this point (\fig{fig:tethered-results}c; $\diamond$). 

Our results reproduced the French-horn conformation for tethered chains, providing confidence that a significant portion of the DNA can be expected to be strongly stretched.
Future research on rotation-induced macromolecular spooling should perform simulations with semiflexible polymer models that represent DNA dynamics at length scales below the Kuhn length in order to explore the stem region of the French-horn conformation in more detail.
To ensure that the strongly stretched stem follows a deterministic screw-like, overlap-free organization, the rotation rate should be slowly increased.
In this way, excluded volume effects act to prevent overlap such that, when the stem forms, it is stretched into a non-overlapping single-file conformation, an important requirement for sequential ordering.
While genetic information in the shofar portion of the DNA remains disordered in space, it is arranged cylindrically in the stem, just as linearly advancing music notes are punched on the cylinder of a music box.

\section{Discussion}
\label{sec:discussion}

Organization of long, single strands of DNA sequentially in space must overcome the high entropy associated with complex, folded polymer conformations.
Rotation-induced macromolecular spooling presents a new single-molecule manipulation concept that explicitly exploits disordering thermal fluctuations through rotation-induced hydrodynamic drag, allowing DNA to be ordered into a curvilinear progression of base-pairs, arranged sequentially on the surface of a rotating microwire.

We observe untethered strands to be highly deformed, and cross streamlines towards the microwire, essentially generating a system capable of studying the single-molecule origin of the Weissenberg effect in rod-climbing of polymeric liquids~\cite{morris2014}.
We have further found enhanced extension to be possible with tethering.
It is compelling that DNA tethered to the rotating microwire takes various previously-unseen conformational states, including the experimentally observed shofar and computationally observed French-horn conformations, at moderate rotations rates. 
The portion of the French-horn conformation that is in the fully extended stem state increases with both rotation rate and contour length.
Longer strands, such as genomic-length DNA, are seen to have a greater portion of their segments in the strongly stretched stem regime of the French-horn conformation. 
In fact, by anchoring one end of a DNA strand to the microwire after utilizing DNA ligase to conjoin a long ``drag-tag'' of additional DNA (that is not of interest) to the free end to act as the unordered shofar-shaped portion of the French-horn conformation, it is possible that future applications could tautly wrap entire DNA lengths into a deterministic single-file and unknotted stem conformation encircling and forming a genomic rim around the microwire. 

Since the single-file stem conformation results from shear drag on the shofar end, rotation-induced macromolecular spooling of very long strands of fragile DNA may face limitations due to shear-induced fragmentation~\cite{Dancis1978,Shui2011,li2017}, overstretching~\cite{Williams2002,bustamante2003,Shokri2008} or tether failing. 
Within the stem, the tension is largest and the rate of tether breakage events may be increased or overstretching may occur at high rotation rates. 
Although we do not observe instances of fragmentation of tethered DNA in the current implementation, we do see tethered strands disappear after repeated spinning, indicating that the tether has failed. 
We speculate that gently ramping up the rotation rate will both reduce sudden fragmentation and allow excluded-volume interactions to reduce instances of entangled conformations or overlaps within the strongly stretched stem. 

Furthermore, a subsequent experimental step could be implemented to immobilize DNA on the microwire, following its ordering on the surface, for storage and readout.
In this way, the complete polymer could be organized into a linear progression of base-pairs that are not susceptible to thermal fluctuations as random coils.
While the relatively short viral genomic DNA considered here ($\sim\SI{e2}{\kilo\basepair}$; $\sim\SI{e1}{\micro\meter}$) may be fully stretched in linear geometries at micron-scales, the distribution of genome sizes varies enormously~\cite{gregory2005}, such that linearly stretching DNA is an impossibility at micron-scales for many genomes --- an impossibility rotation-induced spooling could elegantly overcome through compact curvilinear sequential organization.
The genomic sequence could be stored on the wire for long periods and at high density, in analogy to punching sequential music notes on the surface of a music box cylinder.

\begin{acknowledgments}
\textbf{Acknowledgments:} This work was primarily supported by funding from the NSERC Discovery Program, McGill University and the University of Ottawa. T.N.S. would like to acknowledge fellowship funding from EMBO (ALTF181-2013) and an ERC Advanced Grant MiCE (291234). We thank Ranya Virk for helpful discussions, and Douglas Tree discussions and for providing the Flory exponent for T4-DNA.
\end{acknowledgments}


\begin{thebibliography}{37}%
\makeatletter
\providecommand \@ifxundefined [1]{%
 \@ifx{#1\undefined}
}%
\providecommand \@ifnum [1]{%
 \ifnum #1\expandafter \@firstoftwo
 \else \expandafter \@secondoftwo
 \fi
}%
\providecommand \@ifx [1]{%
 \ifx #1\expandafter \@firstoftwo
 \else \expandafter \@secondoftwo
 \fi
}%
\providecommand \natexlab [1]{#1}%
\providecommand \enquote  [1]{``#1''}%
\providecommand \bibnamefont  [1]{#1}%
\providecommand \bibfnamefont [1]{#1}%
\providecommand \citenamefont [1]{#1}%
\providecommand \href@noop [0]{\@secondoftwo}%
\providecommand \href [0]{\begingroup \@sanitize@url \@href}%
\providecommand \@href[1]{\@@startlink{#1}\@@href}%
\providecommand \@@href[1]{\endgroup#1\@@endlink}%
\providecommand \@sanitize@url [0]{\catcode `\\12\catcode `\$12\catcode
  `\&12\catcode `\#12\catcode `\^12\catcode `\_12\catcode `\%12\relax}%
\providecommand \@@startlink[1]{}%
\providecommand \@@endlink[0]{}%
\providecommand \url  [0]{\begingroup\@sanitize@url \@url }%
\providecommand \@url [1]{\endgroup\@href {#1}{\urlprefix }}%
\providecommand \urlprefix  [0]{URL }%
\providecommand \Eprint [0]{\href }%
\providecommand \doibase [0]{http://dx.doi.org/}%
\providecommand \selectlanguage [0]{\@gobble}%
\providecommand \bibinfo  [0]{\@secondoftwo}%
\providecommand \bibfield  [0]{\@secondoftwo}%
\providecommand \translation [1]{[#1]}%
\providecommand \BibitemOpen [0]{}%
\providecommand \bibitemStop [0]{}%
\providecommand \bibitemNoStop [0]{.\EOS\space}%
\providecommand \EOS [0]{\spacefactor3000\relax}%
\providecommand \BibitemShut  [1]{\csname bibitem#1\endcsname}%
\let\auto@bib@innerbib\@empty
\bibitem [{\citenamefont {Bustamante}\ \emph {et~al.}(2000)\citenamefont
  {Bustamante}, \citenamefont {Smith}, \citenamefont {Liphardt},\ and\
  \citenamefont {Smith}}]{bustamante00}%
  \BibitemOpen
  \bibfield  {author} {\bibinfo {author} {\bibfnamefont {C.}~\bibnamefont
  {Bustamante}}, \bibinfo {author} {\bibfnamefont {S.~B.}\ \bibnamefont
  {Smith}}, \bibinfo {author} {\bibfnamefont {J.}~\bibnamefont {Liphardt}}, \
  and\ \bibinfo {author} {\bibfnamefont {D.}~\bibnamefont {Smith}},\
  }\href@noop {} {\bibfield  {journal} {\bibinfo  {journal} {Current Opinion in
  Structural Biology}\ }\textbf {\bibinfo {volume} {10}},\ \bibinfo {pages}
  {279 } (\bibinfo {year} {2000})}\BibitemShut {NoStop}%
\bibitem [{\citenamefont {Bustamante}\ \emph
  {et~al.}(2003{\natexlab{a}})\citenamefont {Bustamante}, \citenamefont
  {Bryant},\ and\ \citenamefont {Smith}}]{bustamante03}%
  \BibitemOpen
  \bibfield  {author} {\bibinfo {author} {\bibfnamefont {C.}~\bibnamefont
  {Bustamante}}, \bibinfo {author} {\bibfnamefont {Z.}~\bibnamefont {Bryant}},
  \ and\ \bibinfo {author} {\bibfnamefont {S.~B.}\ \bibnamefont {Smith}},\
  }\href@noop {} {\bibfield  {journal} {\bibinfo  {journal} {Nature}\ }\textbf
  {\bibinfo {volume} {421}},\ \bibinfo {pages} {423} (\bibinfo {year}
  {2003}{\natexlab{a}})}\BibitemShut {NoStop}%
\bibitem [{\citenamefont {Renner}\ and\ \citenamefont
  {Doyle}(2015)}]{renner15}%
  \BibitemOpen
  \bibfield  {author} {\bibinfo {author} {\bibfnamefont {C.~B.}\ \bibnamefont
  {Renner}}\ and\ \bibinfo {author} {\bibfnamefont {P.}~\bibnamefont {Doyle}},\
  }\href@noop {} {\bibfield  {journal} {\bibinfo  {journal} {Soft Matter}\ }
  (\bibinfo {year} {2015})}\BibitemShut {NoStop}%
\bibitem [{\citenamefont {Jendrejack}\ \emph {et~al.}(2004)\citenamefont
  {Jendrejack}, \citenamefont {Schwartz}, \citenamefont {de~Pablo},\ and\
  \citenamefont {Graham}}]{jendrejack04}%
  \BibitemOpen
  \bibfield  {author} {\bibinfo {author} {\bibfnamefont {R.~M.}\ \bibnamefont
  {Jendrejack}}, \bibinfo {author} {\bibfnamefont {D.~C.}\ \bibnamefont
  {Schwartz}}, \bibinfo {author} {\bibfnamefont {J.~J.}\ \bibnamefont
  {de~Pablo}}, \ and\ \bibinfo {author} {\bibfnamefont {M.~D.}\ \bibnamefont
  {Graham}},\ }\href@noop {} {\bibfield  {journal} {\bibinfo  {journal} {The
  Journal of Chemical Physics}\ }\textbf {\bibinfo {volume} {120}},\ \bibinfo
  {pages} {2513} (\bibinfo {year} {2004})}\BibitemShut {NoStop}%
\bibitem [{\citenamefont {Usta}\ \emph {et~al.}(2006)\citenamefont {Usta},
  \citenamefont {Butler},\ and\ \citenamefont {Ladd}}]{usta06}%
  \BibitemOpen
  \bibfield  {author} {\bibinfo {author} {\bibfnamefont {O.~B.}\ \bibnamefont
  {Usta}}, \bibinfo {author} {\bibfnamefont {J.~E.}\ \bibnamefont {Butler}}, \
  and\ \bibinfo {author} {\bibfnamefont {A.~J.~C.}\ \bibnamefont {Ladd}},\
  }\href@noop {} {\bibfield  {journal} {\bibinfo  {journal} {Physics of
  Fluids}\ }\textbf {\bibinfo {volume} {18}} (\bibinfo {year}
  {2006})}\BibitemShut {NoStop}%
\bibitem [{\citenamefont {Jo}\ \emph {et~al.}(2009)\citenamefont {Jo},
  \citenamefont {Chen}, \citenamefont {de~Pablo},\ and\ \citenamefont
  {Schwartz}}]{jo09}%
  \BibitemOpen
  \bibfield  {author} {\bibinfo {author} {\bibfnamefont {K.}~\bibnamefont
  {Jo}}, \bibinfo {author} {\bibfnamefont {Y.-L.}\ \bibnamefont {Chen}},
  \bibinfo {author} {\bibfnamefont {J.~J.}\ \bibnamefont {de~Pablo}}, \ and\
  \bibinfo {author} {\bibfnamefont {D.~C.}\ \bibnamefont {Schwartz}},\
  }\href@noop {} {\bibfield  {journal} {\bibinfo  {journal} {Lab on a Chip}\
  }\textbf {\bibinfo {volume} {9}},\ \bibinfo {pages} {2348} (\bibinfo {year}
  {2009})}\BibitemShut {NoStop}%
\bibitem [{\citenamefont {Persson}\ and\ \citenamefont
  {Tegenfeldt}(2010)}]{persson10}%
  \BibitemOpen
  \bibfield  {author} {\bibinfo {author} {\bibfnamefont {F.}~\bibnamefont
  {Persson}}\ and\ \bibinfo {author} {\bibfnamefont {J.~O.}\ \bibnamefont
  {Tegenfeldt}},\ }\href@noop {} {\bibfield  {journal} {\bibinfo  {journal}
  {Chemical Society Reviews}\ }\textbf {\bibinfo {volume} {39}},\ \bibinfo
  {pages} {985} (\bibinfo {year} {2010})}\BibitemShut {NoStop}%
\bibitem [{\citenamefont {Berard}\ \emph {et~al.}(2014)\citenamefont {Berard},
  \citenamefont {Michaud}, \citenamefont {Mahshid}, \citenamefont {Ahamed},
  \citenamefont {McFaul}, \citenamefont {Leith}, \citenamefont
  {B\'{e}rub\'{e}}, \citenamefont {Sladek}, \citenamefont {Reisner},\ and\
  \citenamefont {Leslie}}]{Berard2014}%
  \BibitemOpen
  \bibfield  {author} {\bibinfo {author} {\bibfnamefont {D.~J.}\ \bibnamefont
  {Berard}}, \bibinfo {author} {\bibfnamefont {F.}~\bibnamefont {Michaud}},
  \bibinfo {author} {\bibfnamefont {S.}~\bibnamefont {Mahshid}}, \bibinfo
  {author} {\bibfnamefont {M.~J.}\ \bibnamefont {Ahamed}}, \bibinfo {author}
  {\bibfnamefont {C.~M.~J.}\ \bibnamefont {McFaul}}, \bibinfo {author}
  {\bibfnamefont {J.~S.}\ \bibnamefont {Leith}}, \bibinfo {author}
  {\bibfnamefont {P.}~\bibnamefont {B\'{e}rub\'{e}}}, \bibinfo {author}
  {\bibfnamefont {R.}~\bibnamefont {Sladek}}, \bibinfo {author} {\bibfnamefont
  {W.}~\bibnamefont {Reisner}}, \ and\ \bibinfo {author} {\bibfnamefont
  {S.~R.}\ \bibnamefont {Leslie}},\ }\href@noop {} {\bibfield  {journal}
  {\bibinfo  {journal} {Proceedings of the National Academy of Sciences}\
  }\textbf {\bibinfo {volume} {111}},\ \bibinfo {pages} {13295} (\bibinfo
  {year} {2014})}\BibitemShut {NoStop}%
\bibitem [{\citenamefont {Fyta}(2015)}]{fyta2015}%
  \BibitemOpen
  \bibfield  {author} {\bibinfo {author} {\bibfnamefont {M.}~\bibnamefont
  {Fyta}},\ }\href@noop {} {\bibfield  {journal} {\bibinfo  {journal} {Journal
  of Physics: Condensed Matter}\ }\textbf {\bibinfo {volume} {27}},\ \bibinfo
  {pages} {273101} (\bibinfo {year} {2015})}\BibitemShut {NoStop}%
\bibitem [{\citenamefont {Perkins}\ \emph {et~al.}(1997)\citenamefont
  {Perkins}, \citenamefont {Smith},\ and\ \citenamefont {Chu}}]{perkins97}%
  \BibitemOpen
  \bibfield  {author} {\bibinfo {author} {\bibfnamefont {T.~T.}\ \bibnamefont
  {Perkins}}, \bibinfo {author} {\bibfnamefont {D.~E.}\ \bibnamefont {Smith}},
  \ and\ \bibinfo {author} {\bibfnamefont {S.}~\bibnamefont {Chu}},\
  }\href@noop {} {\bibfield  {journal} {\bibinfo  {journal} {Science}\ }\textbf
  {\bibinfo {volume} {276}},\ \bibinfo {pages} {2016} (\bibinfo {year}
  {1997})}\BibitemShut {NoStop}%
\bibitem [{\citenamefont {Saito}\ \emph {et~al.}(2011)\citenamefont {Saito},
  \citenamefont {Sakaue}, \citenamefont {Kaneko}, \citenamefont {Washizu},\
  and\ \citenamefont {Oana}}]{saito2011}%
  \BibitemOpen
  \bibfield  {author} {\bibinfo {author} {\bibfnamefont {T.}~\bibnamefont
  {Saito}}, \bibinfo {author} {\bibfnamefont {T.}~\bibnamefont {Sakaue}},
  \bibinfo {author} {\bibfnamefont {D.}~\bibnamefont {Kaneko}}, \bibinfo
  {author} {\bibfnamefont {M.}~\bibnamefont {Washizu}}, \ and\ \bibinfo
  {author} {\bibfnamefont {H.}~\bibnamefont {Oana}},\ }\href@noop {} {\bibfield
   {journal} {\bibinfo  {journal} {The Journal of Chemical Physics}\ }\textbf
  {\bibinfo {volume} {135}} (\bibinfo {year} {2011})}\BibitemShut {NoStop}%
\bibitem [{\citenamefont {Ladoux}\ and\ \citenamefont
  {Doyle}(2000)}]{ladoux00}%
  \BibitemOpen
  \bibfield  {author} {\bibinfo {author} {\bibfnamefont {B.}~\bibnamefont
  {Ladoux}}\ and\ \bibinfo {author} {\bibfnamefont {P.~S.}\ \bibnamefont
  {Doyle}},\ }\href@noop {} {\bibfield  {journal} {\bibinfo  {journal} {EPL
  (Europhysics Letters)}\ }\textbf {\bibinfo {volume} {52}},\ \bibinfo {pages}
  {511} (\bibinfo {year} {2000})}\BibitemShut {NoStop}%
\bibitem [{\citenamefont {Gratton}\ and\ \citenamefont
  {Slater}(2005)}]{gratton05}%
  \BibitemOpen
  \bibfield  {author} {\bibinfo {author} {\bibfnamefont {Y.}~\bibnamefont
  {Gratton}}\ and\ \bibinfo {author} {\bibfnamefont {G.~W.}\ \bibnamefont
  {Slater}},\ }\href@noop {} {\bibfield  {journal} {\bibinfo  {journal} {The
  European Physical Journal E}\ }\textbf {\bibinfo {volume} {17}},\ \bibinfo
  {pages} {455} (\bibinfo {year} {2005})}\BibitemShut {NoStop}%
\bibitem [{\citenamefont {Balin}\ \emph {et~al.}()\citenamefont {Balin},
  \citenamefont {Z\"{o}ttl}, \citenamefont {Yeomans},\ and\ \citenamefont
  {Shendruk}}]{Balin2017}%
  \BibitemOpen
  \bibfield  {author} {\bibinfo {author} {\bibfnamefont {A.~K.}\ \bibnamefont
  {Balin}}, \bibinfo {author} {\bibfnamefont {A.}~\bibnamefont {Z\"{o}ttl}},
  \bibinfo {author} {\bibfnamefont {J.~M.}\ \bibnamefont {Yeomans}}, \ and\
  \bibinfo {author} {\bibfnamefont {T.~N.}\ \bibnamefont {Shendruk}},\
  }\href@noop {} {\bibfield  {journal} {\bibinfo  {journal} {Physical Review
  Fluids}\ }}\bibinfo {note} {(submitted)}\BibitemShut {NoStop}%
\bibitem [{\citenamefont {Graham}(2011)}]{graham11}%
  \BibitemOpen
  \bibfield  {author} {\bibinfo {author} {\bibfnamefont {M.~D.}\ \bibnamefont
  {Graham}},\ }\href@noop {} {\bibfield  {journal} {\bibinfo  {journal} {Annual
  Review of Fluid Mechanics}\ }\textbf {\bibinfo {volume} {43}},\ \bibinfo
  {pages} {273} (\bibinfo {year} {2011})}\BibitemShut {NoStop}%
\bibitem [{\citenamefont {G{\"u}nther}\ \emph {et~al.}(2010)\citenamefont
  {G{\"u}nther}, \citenamefont {Mertig},\ and\ \citenamefont
  {Seidel}}]{Gunther10}%
  \BibitemOpen
  \bibfield  {author} {\bibinfo {author} {\bibfnamefont {K.}~\bibnamefont
  {G{\"u}nther}}, \bibinfo {author} {\bibfnamefont {M.}~\bibnamefont {Mertig}},
  \ and\ \bibinfo {author} {\bibfnamefont {R.}~\bibnamefont {Seidel}},\
  }\href@noop {} {\bibfield  {journal} {\bibinfo  {journal} {Nucleic Acids
  Research}\ }\textbf {\bibinfo {volume} {38}},\ \bibinfo {pages} {6526}
  (\bibinfo {year} {2010})}\BibitemShut {NoStop}%
\bibitem [{\citenamefont {Balducci}\ \emph {et~al.}(2006)\citenamefont
  {Balducci}, \citenamefont {Mao}, \citenamefont {Han},\ and\ \citenamefont
  {Doyle}}]{Balducci06}%
  \BibitemOpen
  \bibfield  {author} {\bibinfo {author} {\bibfnamefont {A.}~\bibnamefont
  {Balducci}}, \bibinfo {author} {\bibfnamefont {P.}~\bibnamefont {Mao}},
  \bibinfo {author} {\bibfnamefont {J.}~\bibnamefont {Han}}, \ and\ \bibinfo
  {author} {\bibfnamefont {P.~S.}\ \bibnamefont {Doyle}},\ }\href@noop {}
  {\bibfield  {journal} {\bibinfo  {journal} {Macromolecules}\ }\textbf
  {\bibinfo {volume} {39}},\ \bibinfo {pages} {6273} (\bibinfo {year}
  {2006})}\BibitemShut {NoStop}%
\bibitem [{\citenamefont {Tree}\ \emph {et~al.}(2013)\citenamefont {Tree},
  \citenamefont {Muralidhar}, \citenamefont {Doyle},\ and\ \citenamefont
  {Dorfman}}]{Tree13}%
  \BibitemOpen
  \bibfield  {author} {\bibinfo {author} {\bibfnamefont {D.~R.}\ \bibnamefont
  {Tree}}, \bibinfo {author} {\bibfnamefont {A.}~\bibnamefont {Muralidhar}},
  \bibinfo {author} {\bibfnamefont {P.~S.}\ \bibnamefont {Doyle}}, \ and\
  \bibinfo {author} {\bibfnamefont {K.~D.}\ \bibnamefont {Dorfman}},\
  }\href@noop {} {\bibfield  {journal} {\bibinfo  {journal} {Macromolecules}\
  }\textbf {\bibinfo {volume} {46}},\ \bibinfo {pages} {8369} (\bibinfo {year}
  {2013})}\BibitemShut {NoStop}%
\bibitem [{\citenamefont {Klepinger}\ \emph {et~al.}(2015)\citenamefont
  {Klepinger}, \citenamefont {Greenier},\ and\ \citenamefont
  {Levy}}]{klepinger2015}%
  \BibitemOpen
  \bibfield  {author} {\bibinfo {author} {\bibfnamefont {A.~C.}\ \bibnamefont
  {Klepinger}}, \bibinfo {author} {\bibfnamefont {M.~K.}\ \bibnamefont
  {Greenier}}, \ and\ \bibinfo {author} {\bibfnamefont {S.~L.}\ \bibnamefont
  {Levy}},\ }\href@noop {} {\bibfield  {journal} {\bibinfo  {journal}
  {Macromolecules}\ }\textbf {\bibinfo {volume} {48}},\ \bibinfo {pages} {9007}
  (\bibinfo {year} {2015})}\BibitemShut {NoStop}%
\bibitem [{\citenamefont {Teixeira}\ \emph {et~al.}(2005)\citenamefont
  {Teixeira}, \citenamefont {Babcock}, \citenamefont {Shaqfeh},\ and\
  \citenamefont {Chu}}]{Teixeira05}%
  \BibitemOpen
  \bibfield  {author} {\bibinfo {author} {\bibfnamefont {R.~E.}\ \bibnamefont
  {Teixeira}}, \bibinfo {author} {\bibfnamefont {H.~P.}\ \bibnamefont
  {Babcock}}, \bibinfo {author} {\bibfnamefont {E.~S.~G.}\ \bibnamefont
  {Shaqfeh}}, \ and\ \bibinfo {author} {\bibfnamefont {S.}~\bibnamefont
  {Chu}},\ }\href@noop {} {\bibfield  {journal} {\bibinfo  {journal}
  {Macromolecules}\ }\textbf {\bibinfo {volume} {38}},\ \bibinfo {pages} {581}
  (\bibinfo {year} {2005})}\BibitemShut {NoStop}%
\bibitem [{\citenamefont {Schroeder}\ \emph
  {et~al.}(2005{\natexlab{a}})\citenamefont {Schroeder}, \citenamefont
  {Teixeira}, \citenamefont {Shaqfeh},\ and\ \citenamefont
  {Chu}}]{Schroeder05a}%
  \BibitemOpen
  \bibfield  {author} {\bibinfo {author} {\bibfnamefont {C.~M.}\ \bibnamefont
  {Schroeder}}, \bibinfo {author} {\bibfnamefont {R.~E.}\ \bibnamefont
  {Teixeira}}, \bibinfo {author} {\bibfnamefont {E.~S.~G.}\ \bibnamefont
  {Shaqfeh}}, \ and\ \bibinfo {author} {\bibfnamefont {S.}~\bibnamefont
  {Chu}},\ }\href@noop {} {\bibfield  {journal} {\bibinfo  {journal} {Physical
  Review Letters}\ }\textbf {\bibinfo {volume} {95}},\ \bibinfo {pages}
  {018301} (\bibinfo {year} {2005}{\natexlab{a}})}\BibitemShut {NoStop}%
\bibitem [{\citenamefont {Slater}\ \emph {et~al.}(2009)\citenamefont {Slater},
  \citenamefont {Holm}, \citenamefont {Chubynsky}, \citenamefont {de~Haan},
  \citenamefont {Dub\'{e}}, \citenamefont {Grass}, \citenamefont {Hickey},
  \citenamefont {Kingsburry}, \citenamefont {Sean}, \citenamefont {Shendruk},\
  and\ \citenamefont {Zhan}}]{slater2009modeling}%
  \BibitemOpen
  \bibfield  {author} {\bibinfo {author} {\bibfnamefont {G.~W.}\ \bibnamefont
  {Slater}}, \bibinfo {author} {\bibfnamefont {C.}~\bibnamefont {Holm}},
  \bibinfo {author} {\bibfnamefont {M.~V.}\ \bibnamefont {Chubynsky}}, \bibinfo
  {author} {\bibfnamefont {H.~W.}\ \bibnamefont {de~Haan}}, \bibinfo {author}
  {\bibfnamefont {A.}~\bibnamefont {Dub\'{e}}}, \bibinfo {author}
  {\bibfnamefont {K.}~\bibnamefont {Grass}}, \bibinfo {author} {\bibfnamefont
  {O.~A.}\ \bibnamefont {Hickey}}, \bibinfo {author} {\bibfnamefont
  {C.}~\bibnamefont {Kingsburry}}, \bibinfo {author} {\bibfnamefont
  {D.}~\bibnamefont {Sean}}, \bibinfo {author} {\bibfnamefont {T.~N.}\
  \bibnamefont {Shendruk}}, \ and\ \bibinfo {author} {\bibfnamefont
  {L.}~\bibnamefont {Zhan}},\ }\href@noop {} {\bibfield  {journal} {\bibinfo
  {journal} {Electrophoresis}\ }\textbf {\bibinfo {volume} {30}},\ \bibinfo
  {pages} {792} (\bibinfo {year} {2009})}\BibitemShut {NoStop}%
\bibitem [{\citenamefont {Jiang}\ \emph {et~al.}(2007)\citenamefont {Jiang},
  \citenamefont {Huang}, \citenamefont {Wang},\ and\ \citenamefont
  {Laradji}}]{Jiang07}%
  \BibitemOpen
  \bibfield  {author} {\bibinfo {author} {\bibfnamefont {W.}~\bibnamefont
  {Jiang}}, \bibinfo {author} {\bibfnamefont {J.}~\bibnamefont {Huang}},
  \bibinfo {author} {\bibfnamefont {Y.}~\bibnamefont {Wang}}, \ and\ \bibinfo
  {author} {\bibfnamefont {M.}~\bibnamefont {Laradji}},\ }\href@noop {}
  {\bibfield  {journal} {\bibinfo  {journal} {The Journal of Chemical Physics}\
  }\textbf {\bibinfo {volume} {126}} (\bibinfo {year} {2007})}\BibitemShut
  {NoStop}%
\bibitem [{\citenamefont {Schroeder}\ \emph
  {et~al.}(2005{\natexlab{b}})\citenamefont {Schroeder}, \citenamefont
  {Teixeira}, \citenamefont {Shaqfeh}, ,\ and\ \citenamefont
  {Chu}}]{Schroeder05b}%
  \BibitemOpen
  \bibfield  {author} {\bibinfo {author} {\bibfnamefont {C.~M.}\ \bibnamefont
  {Schroeder}}, \bibinfo {author} {\bibfnamefont {R.~E.}\ \bibnamefont
  {Teixeira}}, \bibinfo {author} {\bibfnamefont {E.~S.~G.}\ \bibnamefont
  {Shaqfeh}}, , \ and\ \bibinfo {author} {\bibfnamefont {S.}~\bibnamefont
  {Chu}},\ }\href@noop {} {\bibfield  {journal} {\bibinfo  {journal}
  {Macromolecules}\ }\textbf {\bibinfo {volume} {38}},\ \bibinfo {pages} {1967}
  (\bibinfo {year} {2005}{\natexlab{b}})}\BibitemShut {NoStop}%
\bibitem [{\citenamefont {Harasim}\ \emph {et~al.}(2013)\citenamefont
  {Harasim}, \citenamefont {Wunderlich}, \citenamefont {Peleg}, \citenamefont
  {Kr\"oger},\ and\ \citenamefont {Bausch}}]{Harasim13}%
  \BibitemOpen
  \bibfield  {author} {\bibinfo {author} {\bibfnamefont {M.}~\bibnamefont
  {Harasim}}, \bibinfo {author} {\bibfnamefont {B.}~\bibnamefont {Wunderlich}},
  \bibinfo {author} {\bibfnamefont {O.}~\bibnamefont {Peleg}}, \bibinfo
  {author} {\bibfnamefont {M.}~\bibnamefont {Kr\"oger}}, \ and\ \bibinfo
  {author} {\bibfnamefont {A.~R.}\ \bibnamefont {Bausch}},\ }\href@noop {}
  {\bibfield  {journal} {\bibinfo  {journal} {Physical Review Letters}\
  }\textbf {\bibinfo {volume} {110}},\ \bibinfo {pages} {108302} (\bibinfo
  {year} {2013})}\BibitemShut {NoStop}%
\bibitem [{\citenamefont {Brochard-Wyart}(1993)}]{brochardwyart93}%
  \BibitemOpen
  \bibfield  {author} {\bibinfo {author} {\bibfnamefont {F.}~\bibnamefont
  {Brochard-Wyart}},\ }\href@noop {} {\bibfield  {journal} {\bibinfo  {journal}
  {EPL (Europhysics Letters)}\ }\textbf {\bibinfo {volume} {23}},\ \bibinfo
  {pages} {105} (\bibinfo {year} {1993})}\BibitemShut {NoStop}%
\bibitem [{\citenamefont {Laleman}\ \emph {et~al.}(2016)\citenamefont
  {Laleman}, \citenamefont {Baiesi}, \citenamefont {Belotserkovskii},
  \citenamefont {Sakaue}, \citenamefont {Walter},\ and\ \citenamefont
  {Carlon}}]{laleman16}%
  \BibitemOpen
  \bibfield  {author} {\bibinfo {author} {\bibfnamefont {M.}~\bibnamefont
  {Laleman}}, \bibinfo {author} {\bibfnamefont {M.}~\bibnamefont {Baiesi}},
  \bibinfo {author} {\bibfnamefont {B.~P.}\ \bibnamefont {Belotserkovskii}},
  \bibinfo {author} {\bibfnamefont {T.}~\bibnamefont {Sakaue}}, \bibinfo
  {author} {\bibfnamefont {J.-C.}\ \bibnamefont {Walter}}, \ and\ \bibinfo
  {author} {\bibfnamefont {E.}~\bibnamefont {Carlon}},\ }\href@noop {}
  {\bibfield  {journal} {\bibinfo  {journal} {Macromolecules}\ }\textbf
  {\bibinfo {volume} {49}},\ \bibinfo {pages} {405} (\bibinfo {year}
  {2016})}\BibitemShut {NoStop}%
\bibitem [{\citenamefont {Buguin}\ and\ \citenamefont
  {Brochard-Wyart}(1996)}]{buguin96}%
  \BibitemOpen
  \bibfield  {author} {\bibinfo {author} {\bibfnamefont {A.}~\bibnamefont
  {Buguin}}\ and\ \bibinfo {author} {\bibfnamefont {F.}~\bibnamefont
  {Brochard-Wyart}},\ }\href@noop {} {\bibfield  {journal} {\bibinfo  {journal}
  {Macromolecules}\ }\textbf {\bibinfo {volume} {29}},\ \bibinfo {pages} {4937}
  (\bibinfo {year} {1996})}\BibitemShut {NoStop}%
\bibitem [{\citenamefont {Slater}\ \emph {et~al.}(2004)\citenamefont {Slater},
  \citenamefont {Gratton}, \citenamefont {Kenward}, \citenamefont {McCormick},\
  and\ \citenamefont {Tessier}}]{slater04}%
  \BibitemOpen
  \bibfield  {author} {\bibinfo {author} {\bibfnamefont {G.~W.}\ \bibnamefont
  {Slater}}, \bibinfo {author} {\bibfnamefont {Y.}~\bibnamefont {Gratton}},
  \bibinfo {author} {\bibfnamefont {M.}~\bibnamefont {Kenward}}, \bibinfo
  {author} {\bibfnamefont {L.}~\bibnamefont {McCormick}}, \ and\ \bibinfo
  {author} {\bibfnamefont {F.}~\bibnamefont {Tessier}},\ }\href@noop {}
  {\bibfield  {journal} {\bibinfo  {journal} {Soft Materials}\ }\textbf
  {\bibinfo {volume} {2}},\ \bibinfo {pages} {155} (\bibinfo {year}
  {2004})}\BibitemShut {NoStop}%
\bibitem [{\citenamefont {Denn}\ and\ \citenamefont
  {Morris}(2014)}]{morris2014}%
  \BibitemOpen
  \bibfield  {author} {\bibinfo {author} {\bibfnamefont {M.~M.}\ \bibnamefont
  {Denn}}\ and\ \bibinfo {author} {\bibfnamefont {J.~F.}\ \bibnamefont
  {Morris}},\ }\href@noop {} {\bibfield  {journal} {\bibinfo  {journal} {Annual
  Review of Chemical and Biomolecular Engineering}\ }\textbf {\bibinfo {volume}
  {5}},\ \bibinfo {pages} {203} (\bibinfo {year} {2014})}\BibitemShut {NoStop}%
\bibitem [{\citenamefont {Dancis}(1978)}]{Dancis1978}%
  \BibitemOpen
  \bibfield  {author} {\bibinfo {author} {\bibfnamefont {B.~M.}\ \bibnamefont
  {Dancis}},\ }\href@noop {} {\bibfield  {journal} {\bibinfo  {journal}
  {Biophysical Journal}\ }\textbf {\bibinfo {volume} {24}},\ \bibinfo {pages}
  {489 } (\bibinfo {year} {1978})}\BibitemShut {NoStop}%
\bibitem [{\citenamefont {Shui}\ \emph {et~al.}(2011)\citenamefont {Shui},
  \citenamefont {Bomer}, \citenamefont {Jin}, \citenamefont {Carlen},\ and\
  \citenamefont {van~den Berg}}]{Shui2011}%
  \BibitemOpen
  \bibfield  {author} {\bibinfo {author} {\bibfnamefont {L.}~\bibnamefont
  {Shui}}, \bibinfo {author} {\bibfnamefont {J.~G.}\ \bibnamefont {Bomer}},
  \bibinfo {author} {\bibfnamefont {M.}~\bibnamefont {Jin}}, \bibinfo {author}
  {\bibfnamefont {E.~T.}\ \bibnamefont {Carlen}}, \ and\ \bibinfo {author}
  {\bibfnamefont {A.}~\bibnamefont {van~den Berg}},\ }\href@noop {} {\bibfield
  {journal} {\bibinfo  {journal} {Nanotechnology}\ }\textbf {\bibinfo {volume}
  {22}},\ \bibinfo {pages} {494013} (\bibinfo {year} {2011})}\BibitemShut
  {NoStop}%
\bibitem [{\citenamefont {Li}\ \emph {et~al.}(2017)\citenamefont {Li},
  \citenamefont {Jin}, \citenamefont {Sun}, \citenamefont {Wang}, \citenamefont
  {Xie}, \citenamefont {Zhou}, \citenamefont {Van~den Berg}, \citenamefont
  {Eijkel},\ and\ \citenamefont {Shui}}]{li2017}%
  \BibitemOpen
  \bibfield  {author} {\bibinfo {author} {\bibfnamefont {L.}~\bibnamefont
  {Li}}, \bibinfo {author} {\bibfnamefont {M.}~\bibnamefont {Jin}}, \bibinfo
  {author} {\bibfnamefont {C.}~\bibnamefont {Sun}}, \bibinfo {author}
  {\bibfnamefont {X.}~\bibnamefont {Wang}}, \bibinfo {author} {\bibfnamefont
  {S.}~\bibnamefont {Xie}}, \bibinfo {author} {\bibfnamefont {G.}~\bibnamefont
  {Zhou}}, \bibinfo {author} {\bibfnamefont {A.}~\bibnamefont {Van~den Berg}},
  \bibinfo {author} {\bibfnamefont {J.~C.~T.}\ \bibnamefont {Eijkel}}, \ and\
  \bibinfo {author} {\bibfnamefont {L.}~\bibnamefont {Shui}},\ }\href@noop {}
  {\bibfield  {journal} {\bibinfo  {journal} {Scientific Reports}\ }\textbf
  {\bibinfo {volume} {7}} (\bibinfo {year} {2017})}\BibitemShut {NoStop}%
\bibitem [{\citenamefont {Williams}\ and\ \citenamefont
  {Rouzina}(2002)}]{Williams2002}%
  \BibitemOpen
  \bibfield  {author} {\bibinfo {author} {\bibfnamefont {M.~C.}\ \bibnamefont
  {Williams}}\ and\ \bibinfo {author} {\bibfnamefont {I.}~\bibnamefont
  {Rouzina}},\ }\href@noop {} {\bibfield  {journal} {\bibinfo  {journal}
  {Current Opinion in Structural Biology}\ }\textbf {\bibinfo {volume} {12}},\
  \bibinfo {pages} {330 } (\bibinfo {year} {2002})}\BibitemShut {NoStop}%
\bibitem [{\citenamefont {Bustamante}\ \emph
  {et~al.}(2003{\natexlab{b}})\citenamefont {Bustamante}, \citenamefont
  {Bryant},\ and\ \citenamefont {Smith}}]{bustamante2003}%
  \BibitemOpen
  \bibfield  {author} {\bibinfo {author} {\bibfnamefont {C.}~\bibnamefont
  {Bustamante}}, \bibinfo {author} {\bibfnamefont {Z.}~\bibnamefont {Bryant}},
  \ and\ \bibinfo {author} {\bibfnamefont {S.~B.}\ \bibnamefont {Smith}},\
  }\href@noop {} {\bibfield  {journal} {\bibinfo  {journal} {Nature}\ }\textbf
  {\bibinfo {volume} {421}},\ \bibinfo {pages} {423} (\bibinfo {year}
  {2003}{\natexlab{b}})}\BibitemShut {NoStop}%
\bibitem [{\citenamefont {Shokri}\ \emph {et~al.}(2008)\citenamefont {Shokri},
  \citenamefont {McCauley}, \citenamefont {Rouzina},\ and\ \citenamefont
  {Williams}}]{Shokri2008}%
  \BibitemOpen
  \bibfield  {author} {\bibinfo {author} {\bibfnamefont {L.}~\bibnamefont
  {Shokri}}, \bibinfo {author} {\bibfnamefont {M.~J.}\ \bibnamefont
  {McCauley}}, \bibinfo {author} {\bibfnamefont {I.}~\bibnamefont {Rouzina}}, \
  and\ \bibinfo {author} {\bibfnamefont {M.~C.}\ \bibnamefont {Williams}},\
  }\href@noop {} {\bibfield  {journal} {\bibinfo  {journal} {Biophysical
  Journal}\ }\textbf {\bibinfo {volume} {95}},\ \bibinfo {pages} {1248 }
  (\bibinfo {year} {2008})}\BibitemShut {NoStop}%
\bibitem [{\citenamefont {Gregory}(2005)}]{gregory2005}%
  \BibitemOpen
  \bibfield  {author} {\bibinfo {author} {\bibfnamefont {T.~R.}\ \bibnamefont
  {Gregory}},\ }\href@noop {} {\bibfield  {journal} {\bibinfo  {journal}
  {Nature Reviews Genetics}\ }\textbf {\bibinfo {volume} {6}},\ \bibinfo
  {pages} {699} (\bibinfo {year} {2005})}\BibitemShut {NoStop}%
\end{thebibliography}
%

\end{document}